\documentclass[twocolumn]{revtex4}
\usepackage{amssymb}
\usepackage{latexsym}
\usepackage{epsfig}

\usepackage{amsmath}
\begin{document}
\title{Tsallis Agegraphic Dark Energy Model}
\author{M. Abdollahi Zadeh$^{1}$\footnote{
m.abdollahizadeh@shirazu.ac.ir}, A.
Sheykhi$^{1,2}$\footnote{asheykhi@shirazu.ac.ir}, H. Moradpour$^{2}$\footnote{h.moradpour@riaam.ac.ir}}
\address{$^1$ Physics Department and Biruni Observatory, College of
Sciences, Shiraz University, Shiraz 71454, Iran\\
$^2$ Research Institute for Astronomy and Astrophysics of Maragha
(RIAAM), P.O. Box 55134-441, Maragha, Iran}

\begin{abstract}
Using the nonextensive Tsallis entropy and the holographic
hypothesis, we propose a new dark energy (DE) model with time
scale as infrared (IR) cutoff. Considering the age of the Universe
as well as the conformal time as IR cutoffs, we investigate the
cosmological consequences of the proposed DE models and study the
evolution of the Universe filled by a pressureless matter and the
obtained DE candidates. We find that although these models can
describe the late time acceleration and the density, deceleration
and the equation of state parameters show satisfactory behavior by
themselves, however, these models are classically unstable unless
the interaction between the two dark sectors of the Universe is
taken into account. In addition, the results of the existence of a
mutual interaction between the cosmos sectors are also addressed.
We find out that the interacting models are stable at the
classical level which is in contrast to the original interacting
agegraphic dark energy models which are classically unstable
\cite{kyoung}.
\end{abstract}
\maketitle

\section{Introduction}
The cosmological constant $\Lambda$ \cite{Padmanabham} is the
simplest approach to DE puzzle, which is responsible for the
current acceleration of the Universe expansion
\cite{Riess,Riess1}, and fills about \%70 of energy content of the
cosmos \cite{Peebles, Ratra, Wetterich}. It may also be described
by modifying general relativity (GR) \cite{SWR,Lombriser}. In
addition, recent observations admit a mutual interaction ($Q$)
between the dark matter (DM) and DE \cite{Amendola,Wang,int,SME}
meaning that their evolution is not independent from each other, a
result decomposes the total energy-momentum conservation law as
\begin{eqnarray}\label{conm}
&&\dot{\rho}_m+3H\rho_m=Q,\\
&&\dot{\rho}_D+3H(1+\omega_D)\rho_D=-Q,\label{conD}
\end{eqnarray}
where $\rho_D$ and $\rho_m$ are the DE and DM energy densities,
respectively. $\omega_D\equiv\frac{p_D}{\rho_D}$, where
$p_D$ denotes the pressure of DE, is also called the state
parameter of DE. Such interaction may solve the coincidence
problem \cite{Salvatelli}, and if $Q<0$ ($Q>0$), then there is an
energy transfer from DM (DE) to DE (DM). See the review
\cite{Wang} for more details and references about the interacting
DE models. Although the $\Lambda$ model is consistent with
observations \cite{Planck}, it suffers some difficulties such as
the fine-tuning and cosmic coincidence problems. These
difficulties motivate physicists to look for other DE candidates.

Agegraphic dark energy (ADE) is an alternative to the $\Lambda$
model based on the uncertainty relation of quantum mechanics
\cite{Cai}. It was argued that in Minkowskian spacetime, the
uncertainty in time $t$ is $\delta{t}=\beta t_{p}^{2/3}t^{1/3}$
where $\beta$ is a dimensionless constant of order unity and
$t_{p}$ denotes the reduced Plank time \cite{Cai}. Due to some
difficulties of the original ADE \cite{Cai}, Wei and Cai proposed
a new ADE model \cite{Wei}, in which the conformal time $\eta$,
defined as $dt=a d\eta$, where $t$ is the cosmic time, is taken
into account instead of the universe age. It is worthwhile
mentioning that the conformal time $\eta$ satisfies the
$ds^2=dt^2-a^2d\mathrm{x^2}=a^2(d\eta^2-d\mathrm{x^2})$ relation
in the FRW universe. Since the entropy relation has a crucial role
in this approach \cite{Cai,Wei,Wei2}, each modification to the
system entropy may change the ADE model. The ADE models  have been
investigated widely in the literatures \cite{ADE,shey1}.

It was first pointed out by Gibbs \cite{Gibbs} that systems with a
long range interaction, such as gravitational systems, do not
necessarily obey the Boltzmann-Gibbs (BG) theory, and indeed these
systems can violate the extensivity constraint of the
Boltzmann-Gibbs entropy. Based on the Gibbs arguments, in $1988$
Tsallis \cite{Tsallis} introduced a statistical description for
the non-extensive systems which leads to a new entropy-area
relation for horizons \cite{Tsallis1}. According to Tsallis, the
entropy associated with the black hole is written as $ S_{\delta}=
\gamma A^{\delta}$, where $\gamma$ is an unknown constant and
$\delta$ denotes the non-extensive parameter. Applying this
non-extensive entropy relation to the apparent horizon of FRW
universe, and using the holographic dark energy hypothesis, a new
holographic DE model was proposed with energy density \cite{THDE},
\begin{eqnarray}\label{Trho}
\rho_D=BL^{2\delta-4},
\end{eqnarray}
where $B$ is an unknown parameter and $L$ is the IR cutoff
\cite{Tsallis1,THDE,THDE1,THDE2,THDE3,THDE4,THDE5,THDE6,THDE7}. More
works in which various non-extensive entropies have been used to
study the cosmic evolution can also be found in
\cite{TDE,SM,Ren,prd,epjc}. Here, we are going to use the
nonextensive Tsallis entropy \cite{Tsallis1} to build two Tsallis
ADE (TADE) models by using the age of the Universe and the
conformal time as the IR cutoffs, and study their effects on the
evolution of the Universe.

The organization of this paper is as follows. In the next section,
we address TADE model in which the age of the Universe is used as
the IR cutoff and study the evolution of the cosmos, whenever
there is no interaction between the two dark sectors. In addition,
the results of considering a mutual interaction between the dark
sectors of cosmos are also investigated. Considering the conformal
time instead of the universe age, we introduce a new ADE model and
study the cosmic evolution in both interacting and non-interacting
FRW universes in the third section. The last section is devoted to
a summary and concluding remarks.
\section{Tsallis Agegraphic dark energy (TADE) model}
Considering the age of the Universe as IR cutoff, which is defined
as
\begin{equation}
T=\int_0^a{dt}=\int_0^a{\frac{da}{Ha}},
\end{equation}
\noindent where $a$ and and $H=\dot{a} / a$  are the scale factor
and the Hubble parameter, respectively,  one can use
Eq.~(\ref{Trho}) to write the energy density of TADE as
\begin{eqnarray}\label{Tage}
\rho_D=BT^{2\delta-4},
\end{eqnarray}
\noindent recovering the primary ADE model of Cai \cite{Cai} at the limit of $\delta=1$. The first Friedmann equation of a flat FRW universe
filled by a pressureless fluid $\rho_m$ and TADE ($\rho_D$) is
written as
\begin{equation}\label{Friedmann}
H^2=\frac{1}{3m_p^2}(\rho_m+\rho_D),
\end{equation}
\noindent which can also be rewritten as
\begin{equation}\label{Fri2}
\Omega_m+\Omega_D=1,
\end{equation}
\noindent by defining the fractional energy densities
\begin{equation}\label{Omega}
\Omega_m=\frac{\rho_m}{3m_p^2H^2},\
 \   \  ~~\Omega_D=\frac{\rho_D}{3m_p^2H^2}.
\end{equation}
\noindent Finally, we easily get
\begin{equation}\label{r}
r =\frac{\Omega_m}{\Omega_D}=-1+\frac{1}{\Omega_D},
\end{equation}
\noindent for the energy densities ratio. As it is apparent from Eq.~(\ref{Omega}), using the observational values of $H$, and the fractional energy densities, one may find primary estimations for the allowed intervals of $B$ and $\delta$ satisfying observations. The more certain results are achievable by employing observational outcomes on other cosmic parameters such as the deceleration parameter and etc. In the following, since we are eager to show the power of this model, considering $\Omega^{0}_D=0.73$ and $H(a=1)=67$ as the current values of these parameters \cite{roos}, we only choose some values of the system parameters such as $\delta$ and $B$ producing distinctive behaviors.
\subsection{Noninteracting case ($Q=0$)}
Inserting the time derivative of Eq.~(\ref{Tage}) in the
conservation equation~(\ref{conD}), one obtains
\begin{eqnarray}\label{EoSa}
\omega_D=-1-\frac{2\delta-4}{3T H},
\end{eqnarray}
\noindent where
\begin{equation}
T=\left(\frac{3 H^2 \Omega_D}{B}\right)^{\frac{1}{2\delta-4}}.
\end{equation}

\begin{figure}[htp]
\begin{center}
\includegraphics[width=8cm]{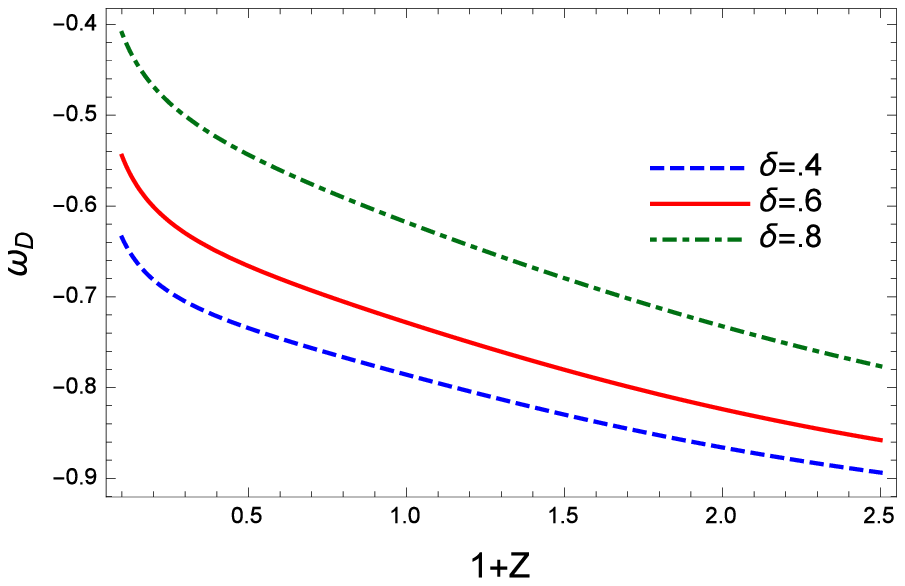}
\includegraphics[width=8cm]{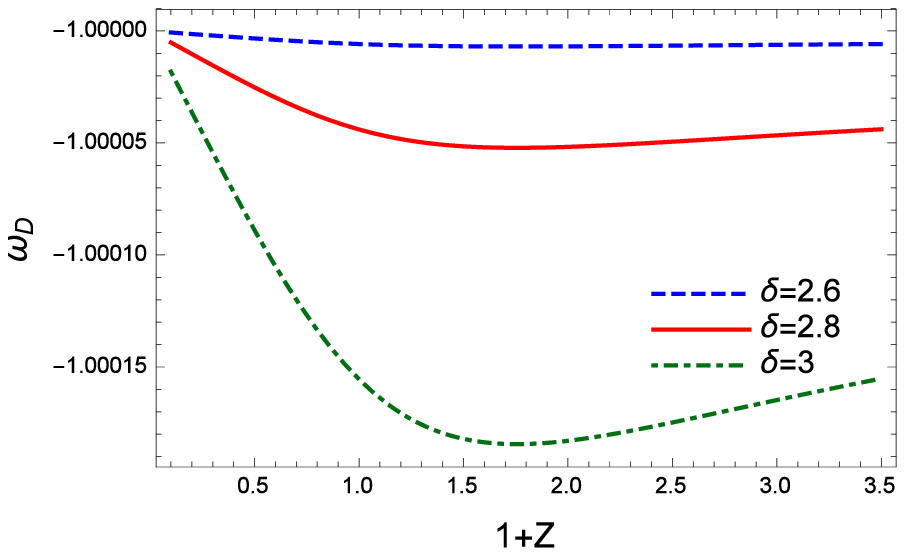}
\caption{Evolution of $\omega_D$ versus redshift parameter $z$ for
non-interacting TADE. Here, we
have taken $\Omega^{0}_D=0.73$, $B=2.4$ and $H(a=1)=67$.
}\label{w-z}
\end{center}
\end{figure}
Additionally, by combining the time derivative of
Eq.~(\ref{Friedmann}) and using Eqs.~(\ref{conm}) and
(\ref{conD}), we reach
\begin{equation}\label{agedotH}
\frac{\dot{H}}{H^2}=-\frac{3}{2}(1-\Omega_D)+\frac{(\delta-2)\Omega_D}{T
H},
\end{equation}
\noindent which can also lead to
\begin{equation}\label{qage}
q\equiv-1-\frac{\dot{H}}{H^2}=-\frac{1}{3}-\frac{3\Omega_D}{2}-\frac{(\delta-2)\Omega_D}{TH},
\end{equation}
\noindent for the deceleration parameter.
\begin{figure}[htp]
\begin{center}
\includegraphics[width=8cm]{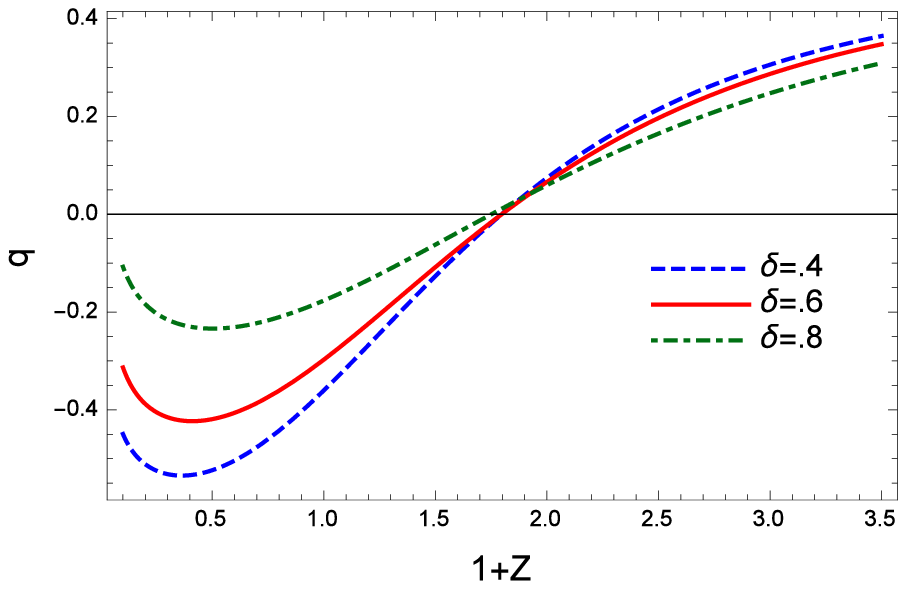}
\includegraphics[width=8cm]{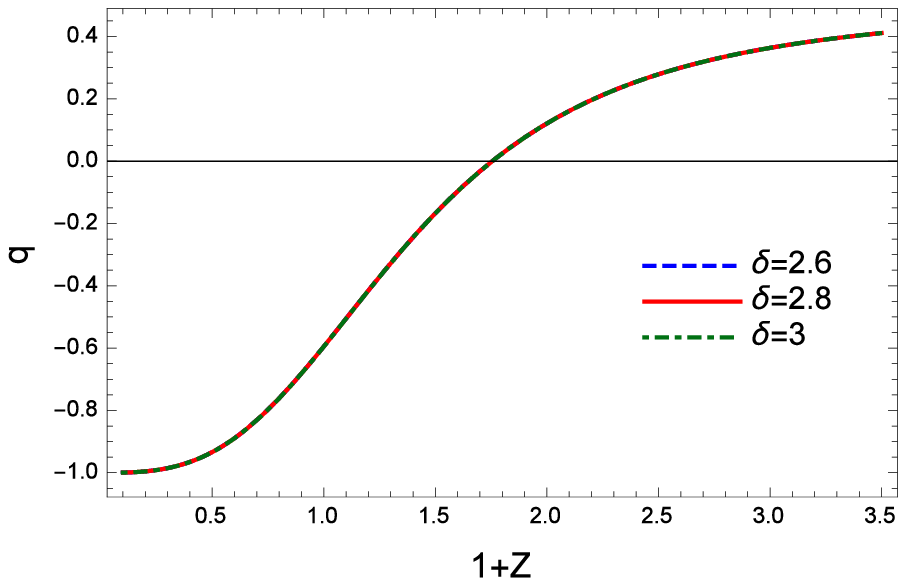}
\caption{Evolution of $q$ versus redshift parameter $z$ for
non-interacting TADE. Here, we
have taken $\Omega^{0}_D=0.73$, $B=2.4$ and $H(a=1)=67$.
}\label{q-z}
\end{center}
\end{figure}
\noindent It is also a matter of calculation to show
\begin{equation}\label{ageOmega}
\dot{\Omega}_D=\frac{(2\delta-4)\Omega_D}{T}+2\Omega_D H(1+q),
\end{equation}
\noindent where dot denotes the derivative with respect to the
cosmic time.
\begin{figure}[htp]
\begin{center}
\includegraphics[width=8cm]{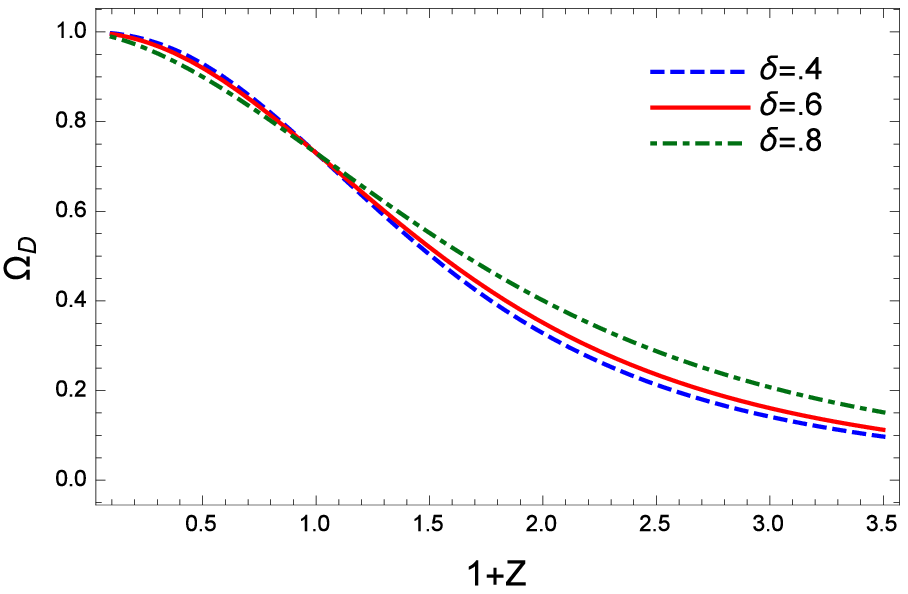}
\includegraphics[width=8cm]{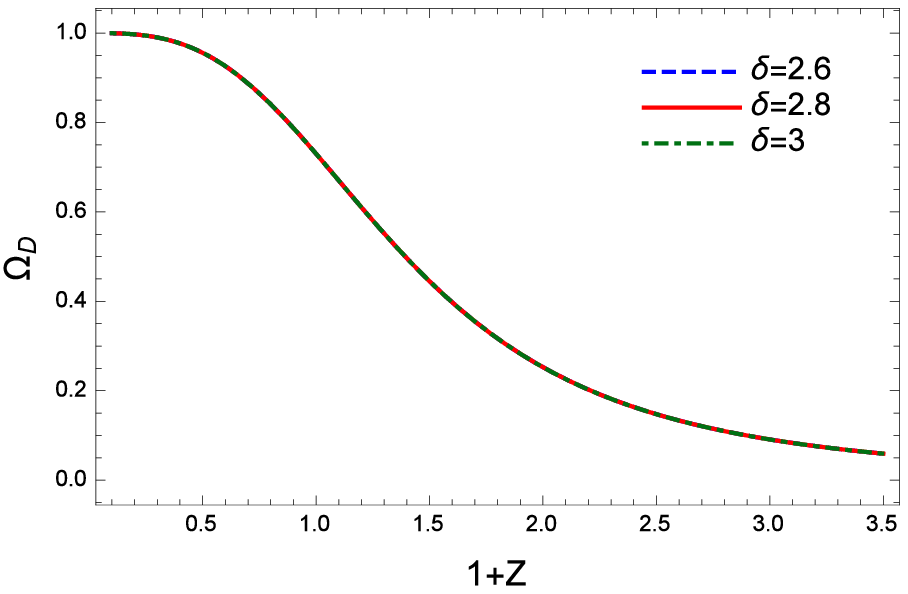}
\caption{Evolution of $\Omega_D$ versus redshift parameter $z$ for
non-interacting TADE. Here, we
have taken $\Omega^{0}_D=0.73$, $B=2.4$ and $H(a=1)=67$.
}\label{o-z}
\end{center}
\end{figure}
In order to study the effects of perturbations on the classical
stability of the model, the squared of the sound speed
($v_{s}^{2}$) should be evaluated,
\begin{equation}\label{vs}
v_{s}^{2}=\frac{dP_D}{d\rho_D}=\frac{\dot{P}_D}{\dot{\rho}_D}=\dfrac{\rho_{D}}{\dot{\rho}_{D}}
\dot{\omega}_{D}+\omega_{D},
\end{equation}
\noindent which finally leads to
\begin{eqnarray}\label{vsage}
&&v_{s}^{2}=\frac{\Omega_D-3}{2}\\&&+\frac{3^{\frac{1}{4-2
\delta}}(H^2\Omega_D B^{-1})^{\frac{1}{4-2\delta}}(5-2\delta+(\delta-2)\Omega_D)}{3H},\nonumber
\end{eqnarray}
\noindent for the non-interacting case.
\begin{figure}[htp]
\begin{center}
\includegraphics[width=8cm]{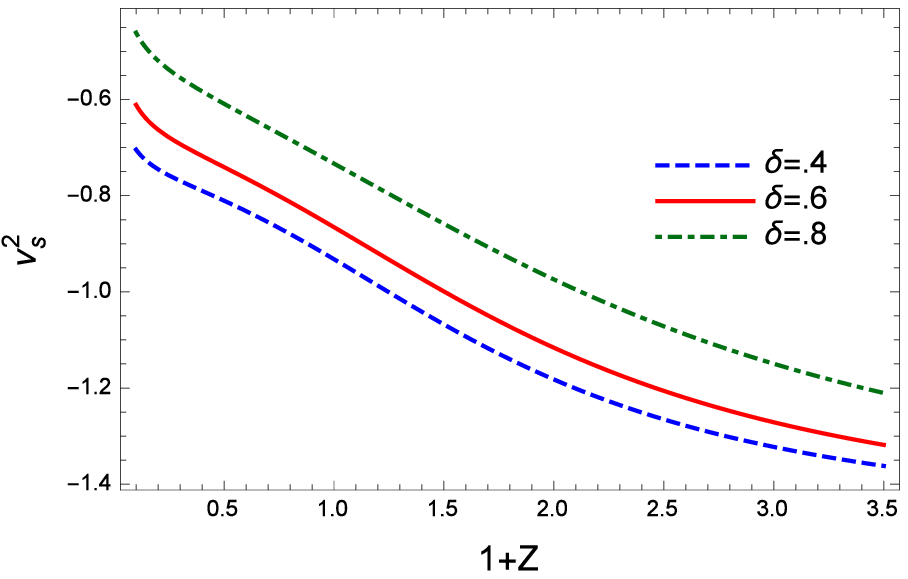}
\includegraphics[width=8cm]{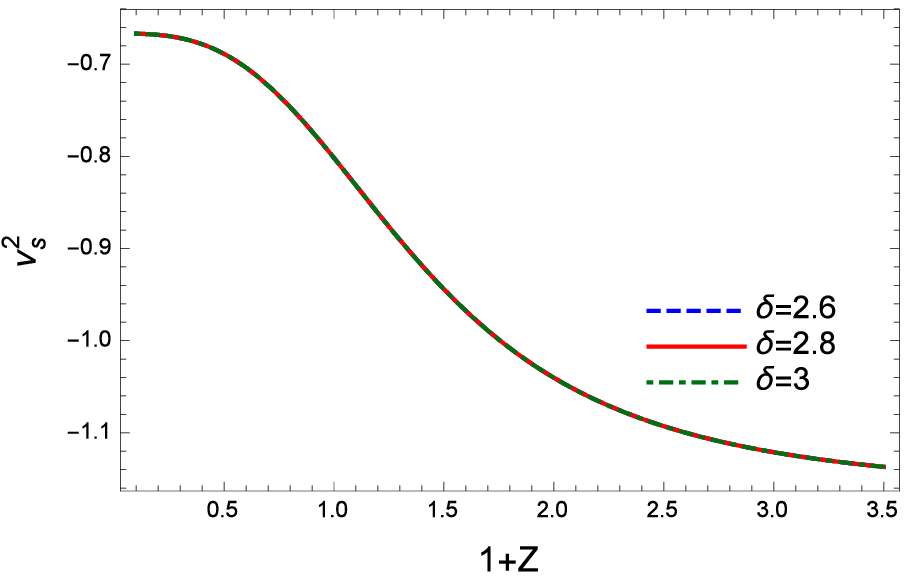}
\caption{Evolution of ${v}^{2}_{s}$ versus redshift parameter $z$ for
non-interacting TADE. Here, we have taken $\Omega^{0}_D=0.73$, $B=2.4$ and $H(a=1)=67$.
}\label{s-z}
\end{center}
\end{figure}
The evolution of the system parameters are plotted in
Figs.~\ref{w-z}-\ref{s-z}. It is apparent that the model is
classically unstable (${v}^{2}_{s}<0$). Moreover, it is apparent
that there are some values of $\delta$ for which $\Omega_D$ and
$q$ can show satisfactory behavior by themselves. The $\delta=2.6$
case is interesting, because, in addition to $\Omega_D$ and $q$,
it leads to suitable behavior for the state parameter
($\omega_D\simeq-1$) during the cosmic evolution.
\subsection{Interacting case ($Q\neq0$)}
\begin{figure}[htp]
\begin{center}
\includegraphics[width=8cm]{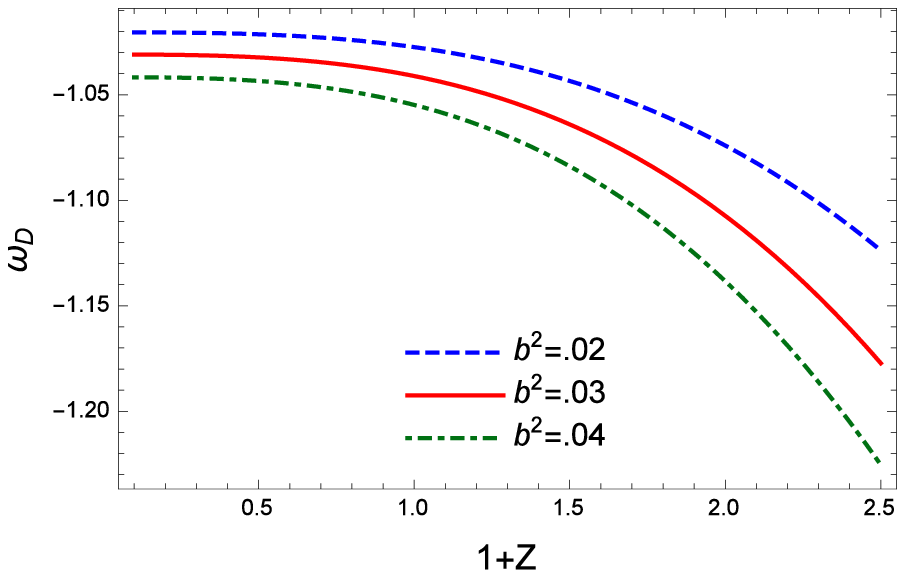}
\caption{Evolution of $\omega_D$ versus redshift parameter $z$ for
interacting TADE. Here, we
have taken $\Omega^{0}_D=0.73$, $B=2.4$, $\delta=2.6$ and $H(a=1)=67$.
}\label{w-z1}
\end{center}
\end{figure}

\begin{figure}[htp]
\begin{center}
\includegraphics[width=8cm]{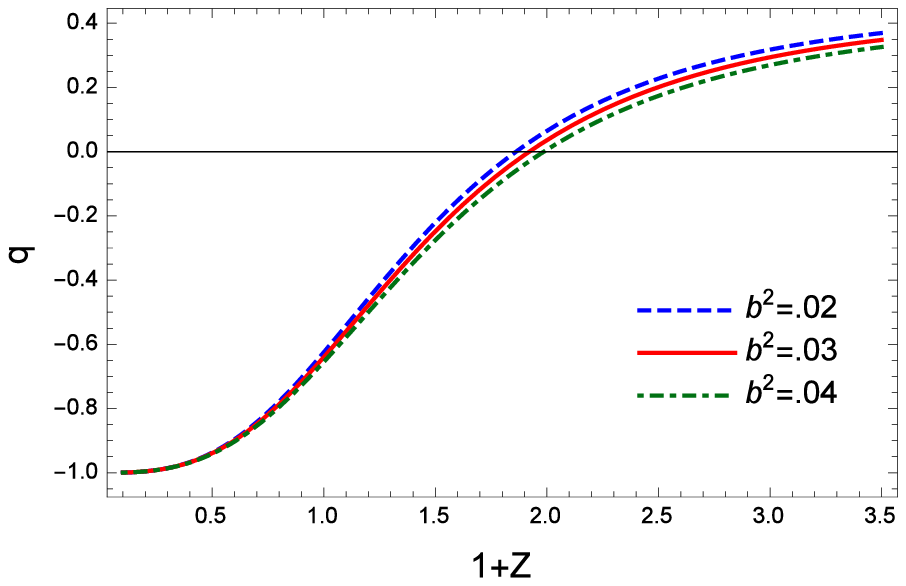}
\caption{Evolution of $q$ versus redshift parameter $z$ for
interacting TADE. Here, we
have taken $\Omega^{0}_D=0.73$, $B=2.4$, $\delta=2.6$ and $H(a=1)=67$.
}\label{q-z1}
\end{center}
\end{figure}

\begin{figure}[htp]
\begin{center}
\includegraphics[width=8cm]{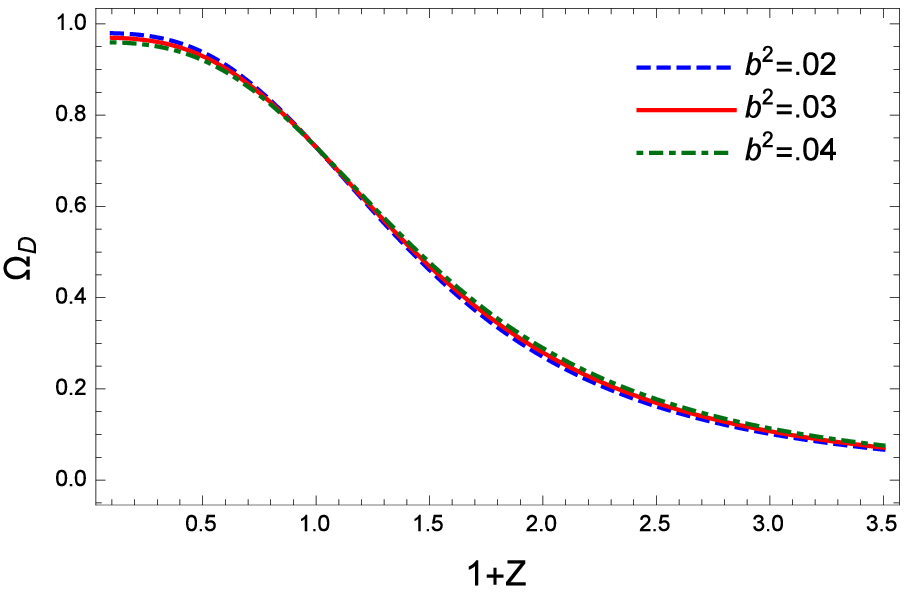}
\caption{Evolution of $\Omega_D$ versus redshift parameter $z$ for
interacting TADE. Here, we
have taken $\Omega^{0}_D=0.73$, $B=2.4$, $\delta=2.6$ and $H(a=1)=67$.
}\label{o-z1}
\end{center}
\end{figure}

\begin{figure}[htp]
\begin{center}
\includegraphics[width=8cm]{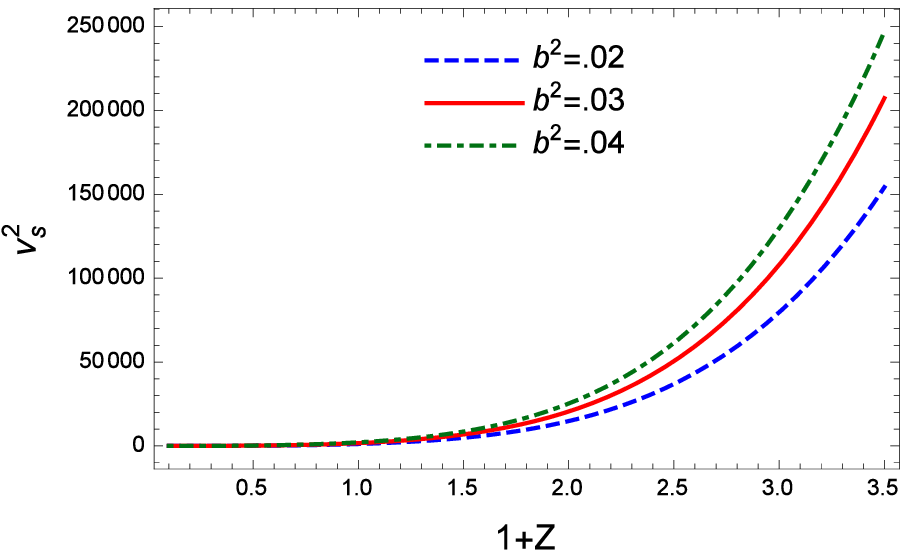}
\caption{Evolution of ${v}^{2}_{s}$ versus redshift parameter $z$ for
interacting TADE. Here, we
have taken $\Omega^{0}_D=0.73$, $B=2.4$, $\delta=2.6$ and $H(a=1)=67$.
}\label{s-z1}
\end{center}
\end{figure}

\begin{figure}[htp]
\begin{center}
\includegraphics[width=8cm]{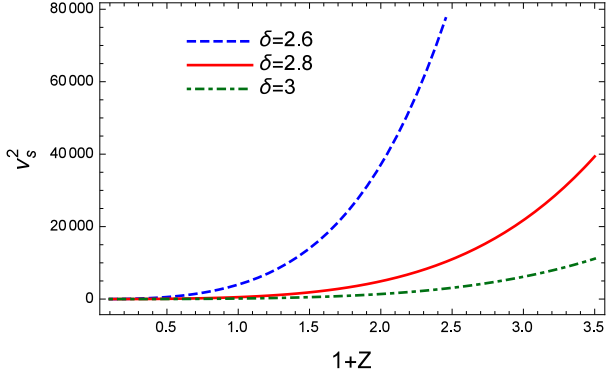}
\caption{Evolution of ${v}^{2}_{s}$ versus redshift parameter $z$ for
interacting TADE. Here, we
have taken $\Omega^{0}_D=0.73$, $B=2.4$, $b^2=.01$ and $H(a=1)=67$.
}\label{ss-z1}
\end{center}
\end{figure}

As mentioned earlier, recent observations indicate that the
evolution of DM and DE is not independent, a key to solve the
coincidence problem \cite{Salvatelli}. Here, the
$Q=3b^2H(\rho_D+\rho_m)$ mutual interaction between the dark
sectors of cosmos \cite{int} is assumed to get the expressions for
deceleration parameter, the equation of state, the evolution of
density parameter, and also $v_{s}^{2}$  as
\begin{eqnarray}\label{qage1}
&&q=-\frac{1}{3}-\frac{3b^2}{2}-\frac{3\Omega_D}{2}-\frac{(\delta-2)\Omega_D}{TH},\nonumber\\ &&\omega_D=-1-\frac{b^2}{\Omega_D}
-\frac{2\delta-4}{3 TH},\nonumber\\ &&\dot{\Omega}_D=\frac{(2\delta-4)\Omega_D}{T}+2\Omega_D H(1+q),\nonumber\\ && v_{s}^{2}
=-\frac{3+b^2-\Omega_D}{2}\\&&-\frac{3^\frac{-7+4\delta}{-4+2\delta}b^2H(H^2\Omega_D B^{-1})^{\frac{1}{-4+2\delta}}
(-1+b^2+\Omega_D)}{6(\delta-2)\Omega_D}\nonumber\\&&
-\frac{3^\frac{-3+2\delta}{4-2\delta}(H^2\Omega_D B^{-1})^{\frac{1}{4-2\delta}}(5-2\delta+(\delta-2)\Omega_D)}{H},\nonumber
\end{eqnarray}
\noindent plotted in Figs.~\ref{w-z1}-\ref{ss-z1} which show
satisfactory behaviors for $\Omega_D$ and $q$. As it is apparent,
this case is classically stable, a behavior unlike those of THDE
\cite{THDE} and the non-interacting case, and moreover, $\omega_D$
acted as that of the phantom sources in past.

\section{New Tsallis agegraphic dark energy model (NTADE)}
Due to some problems of the original ADE \cite{Cai}, a new ADE was
proposed by Wei and Cai \cite{Wei}, in which the conformal time
$\eta$ is used as the IR cutoff instead of the age of the
Universe. The conformal time is defined as $dt=a d\eta$ leading to
$\dot{\eta}=1/a$ and thus
\begin{equation}
\eta=\int_0^a{\frac{da}{Ha^2}}.
\end{equation}
In this manner, using Eq.(\ref{Trho}), the energy density of NTADE is written as
\begin{eqnarray}\label{Tnage}
\rho_D=B{\eta}^{2\delta-4}.
\end{eqnarray}
\subsection{Non-interacting case}
\begin{figure}[htp]
\begin{center}
\includegraphics[width=8cm]{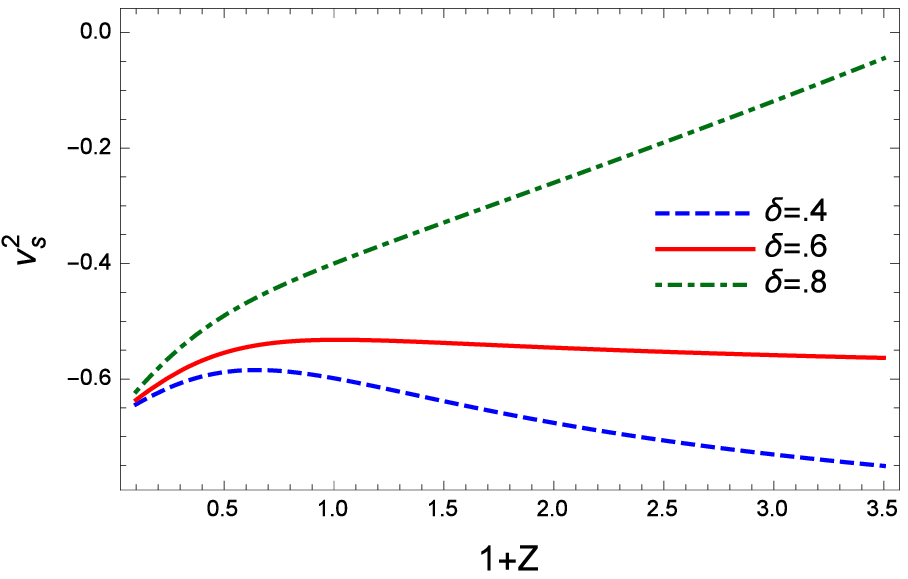}
\includegraphics[width=8cm]{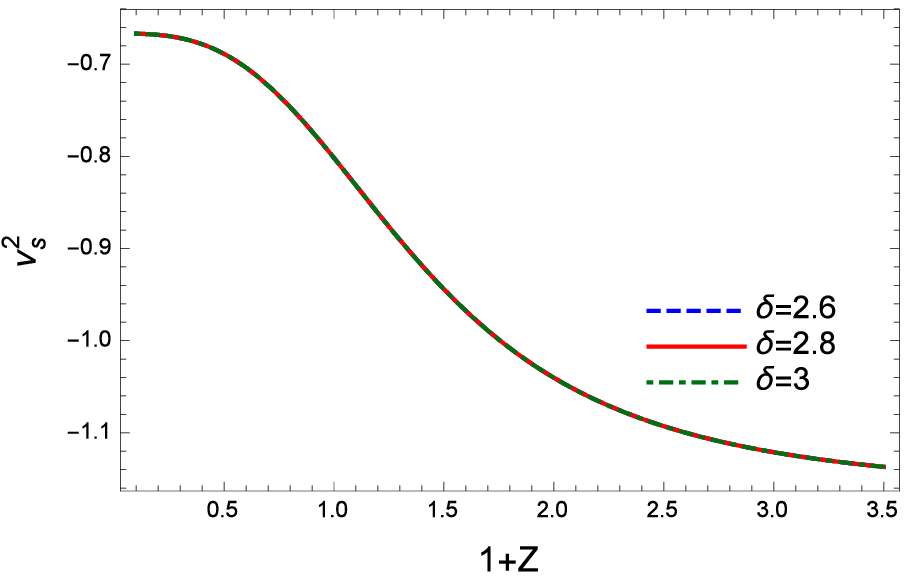}
\caption{Evolution of ${v}^{2}_{s}$ versus redshift parameter $z$ for
non-interacting NTADE. Here, we
have taken $\Omega^{0}_D=0.73$, $B=2.4$ and $H(a=1)=67$.
}\label{sn-z}
\end{center}
\end{figure}

Whenever there is no interaction between the dark sectors of
cosmos ($Q=0$), one can insert Eq.(\ref{Tnage}) and its time
derivative into Eq.(\ref{conD}) to reach
\begin{eqnarray}\label{EoSna}
\omega_D=-1-\frac{2\delta-4}{3a\eta H},
\end{eqnarray}
where $\eta=(\frac{3 H^2 \Omega_D}{B})^{\frac{1}{2\delta-4}}$.
\begin{figure}[htp]
\begin{center}
\includegraphics[width=8cm]{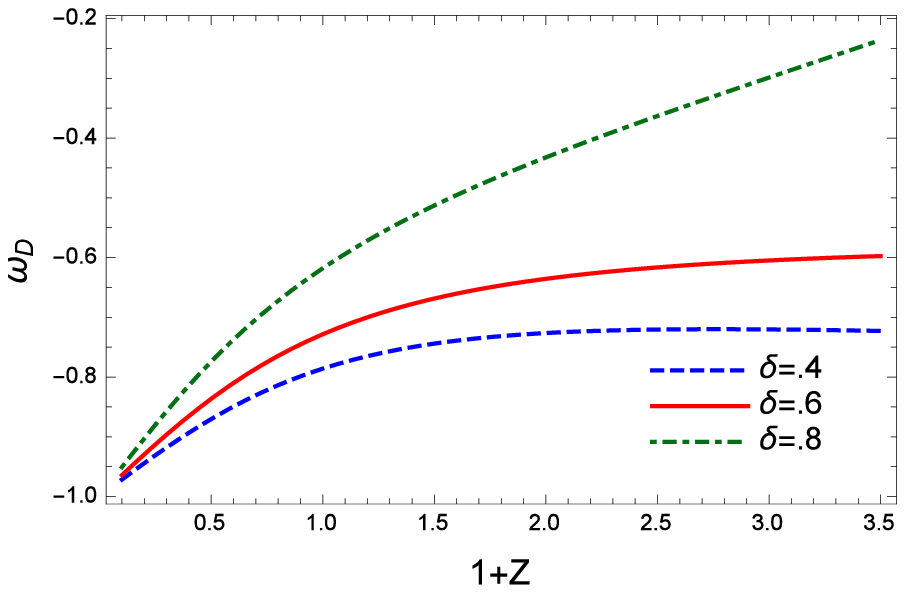}
\includegraphics[width=8cm]{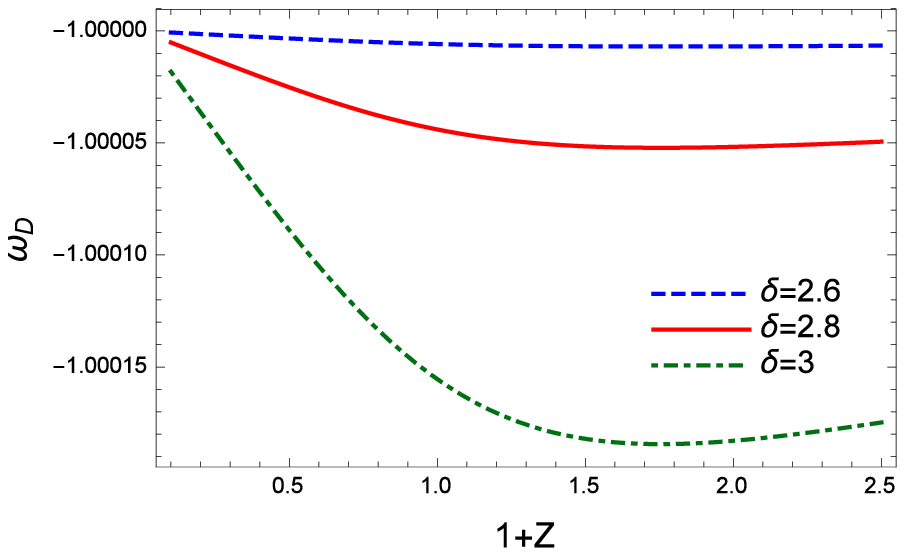}
\caption{Evolution of $\omega_D$ versus redshift parameter $z$ for
non-interacting NTADE. Here, we
have taken $\Omega^{0}_D=0.73$, $B=2.4$ and $H(a=1)=67$.
}\label{wn-z}
\end{center}
\end{figure}
\begin{figure}[htp]
\begin{center}
\includegraphics[width=8cm]{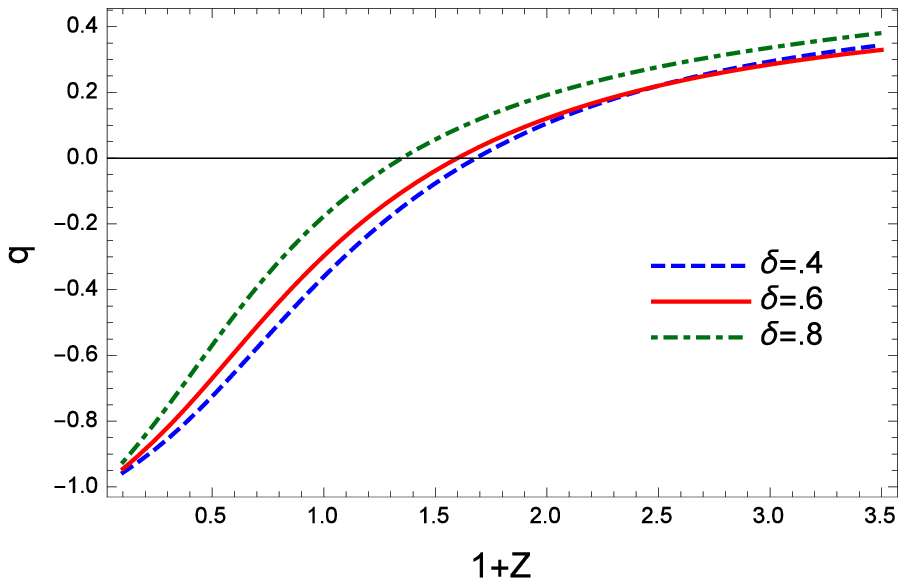}
\includegraphics[width=8cm]{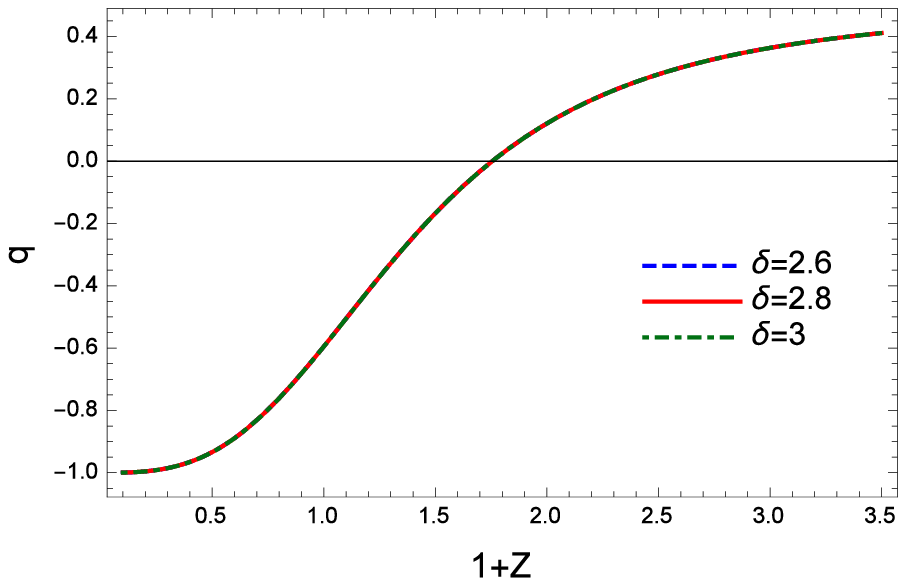}
\caption{Evolution of $q$ versus redshift parameter $z$ for
non-interacting NTADE. Here, we
have taken $\Omega^{0}_D=0.73$, $B=2.4$ and $H(a=1)=67$.
}\label{qn-z}
\end{center}
\end{figure}
Differentiating Eq.(\ref{Friedmann}) and using Eqs. (\ref{Tnage})
and (\ref{conm}), we arrive at
\begin{equation}\label{qnage}
q=-\frac{1}{3}-\frac{3\Omega_D}{2}-\frac{(\delta-2)\Omega_D}{a\eta H},
\end{equation}
for the deceleration parameter. In addition, it is a matter of
calculation to use Eqs. (\ref{Omega}) and~(\ref{Tnage}) in order
to show that
\begin{equation}\label{nageOmega}
\dot{\Omega}_D=\frac{(2\delta-4)\Omega_D}{a\eta}+2\Omega_D H(1+q).
\end{equation}
\begin{figure}[htp]
\begin{center}
\includegraphics[width=8cm]{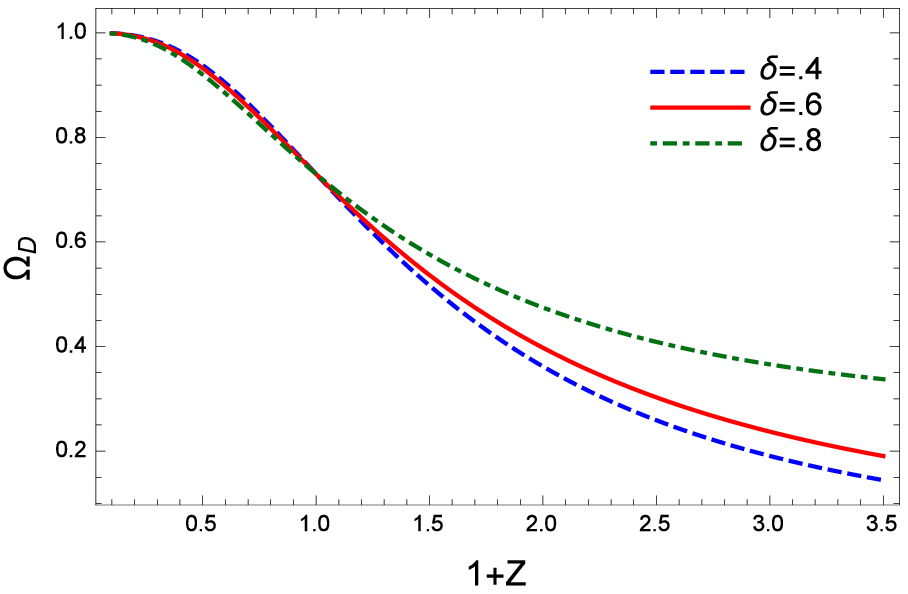}
\includegraphics[width=8cm]{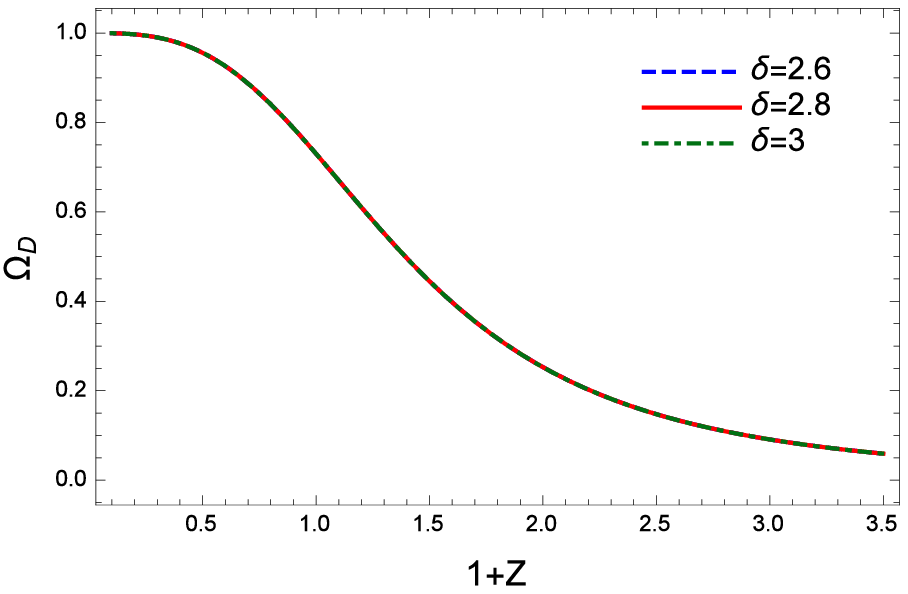}
\caption{Evolution of $\Omega_D$ versus redshift parameter $z$ for
non-interacting NTADE. Here, we
have taken $\Omega^{0}_D=0.73$, $B=2.4$ and $H(a=1)=67$.
}\label{on-z}
\end{center}
\end{figure}
Finally, the squared of the sound speed is find out as
\begin{eqnarray}\label{vsnage}
&&v_{s}^{2}=\frac{3\Omega_D-7}{6}+\\&&\frac{3^{\frac{1}{4-2
\delta}}(H^2\Omega_D B^{-1})^{\frac{1}{4-2\delta}}(5-2\delta+(\delta-2)\Omega_D)}{3H}.\nonumber
\end{eqnarray}
The evolution of the system parameters has been depicted in
Figs.~\ref{wn-z}-\ref{sn-z} claiming that although $\Omega_D$,
$\omega_D$ and $q$ can show acceptable behavior by themselves
during the cosmic evolution, the model is classically unstable
($v_{s}^{2}<0$).
\subsection{Interacting case}
\begin{figure}[htp]
\begin{center}
\includegraphics[width=8cm]{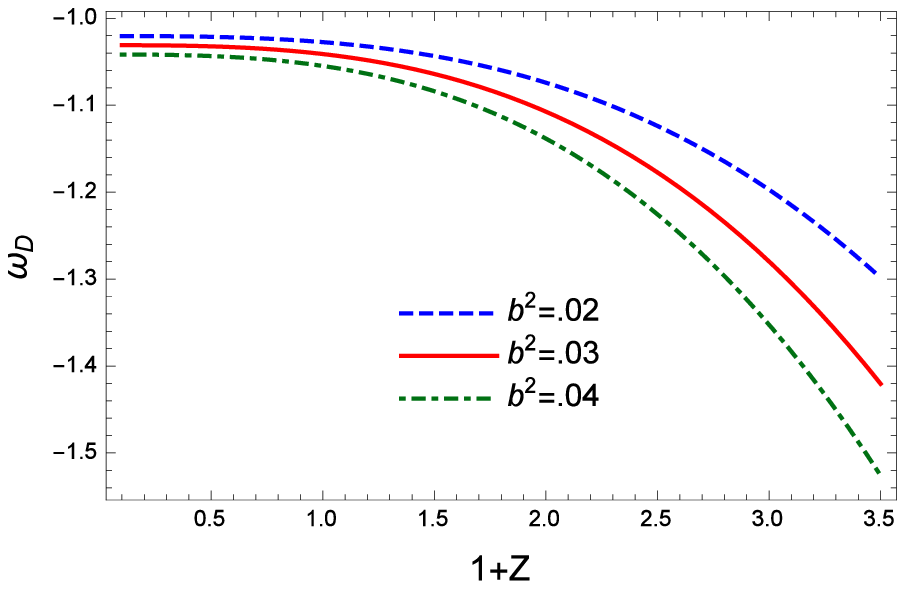}
\caption{Evolution of $\omega_D$ versus redshift parameter $z$ for
interacting NTADE. Here, we
have taken $\Omega^{0}_D=0.73$, $B=2.4$, $\delta=2.6$ and $H(a=1)=67$.
}\label{wn-z1}
\end{center}
\end{figure}

\begin{figure}[htp]
\begin{center}
\includegraphics[width=8cm]{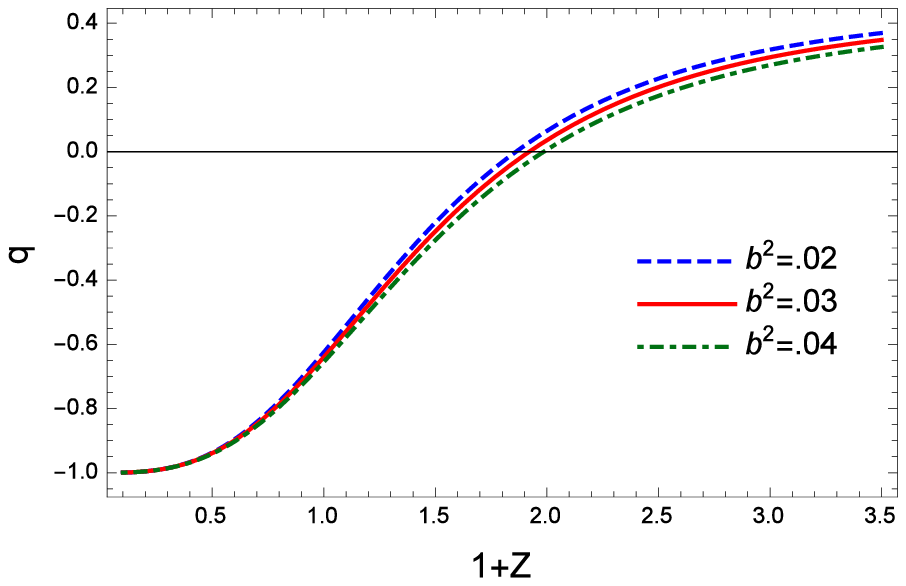}
\caption{Evolution of $q$ versus redshift parameter $z$ for
interacting NTADE. Here, we
have taken $\Omega^{0}_D=0.73$, $B=2.4$, $\delta=2.6$ and $H(a=1)=67$.
}\label{qn-z1}
\end{center}
\end{figure}

\begin{figure}[htp]
\begin{center}
\includegraphics[width=8cm]{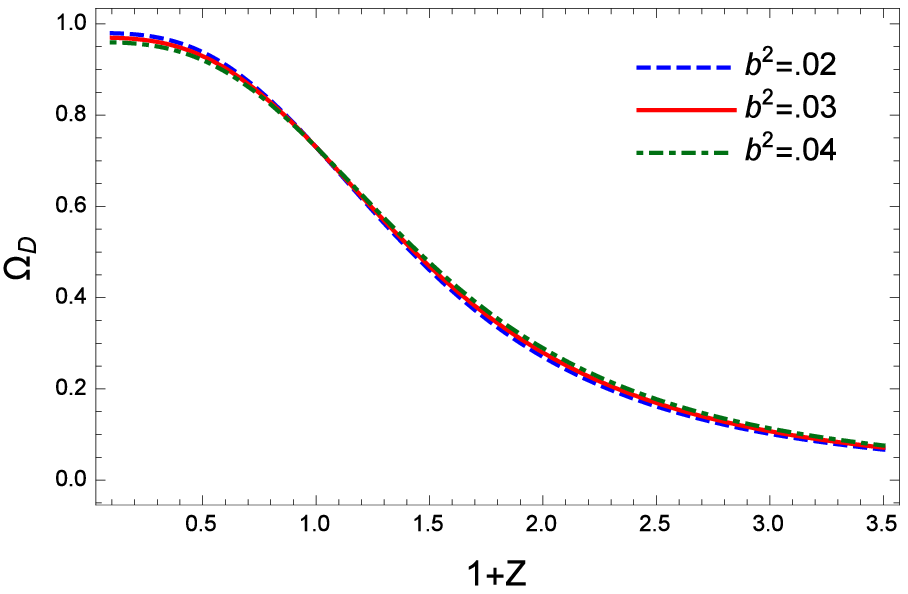}
\caption{Evolution of $\Omega_D$ versus redshift parameter $z$ for
interacting NTADE. Here, we
have taken $\Omega^{0}_D=0.73$, $B=2.4$, $\delta=2.6$ and $H(a=1)=67$.
}\label{on-z1}
\end{center}
\end{figure}

\begin{figure}[htp]
\begin{center}
\includegraphics[width=8cm]{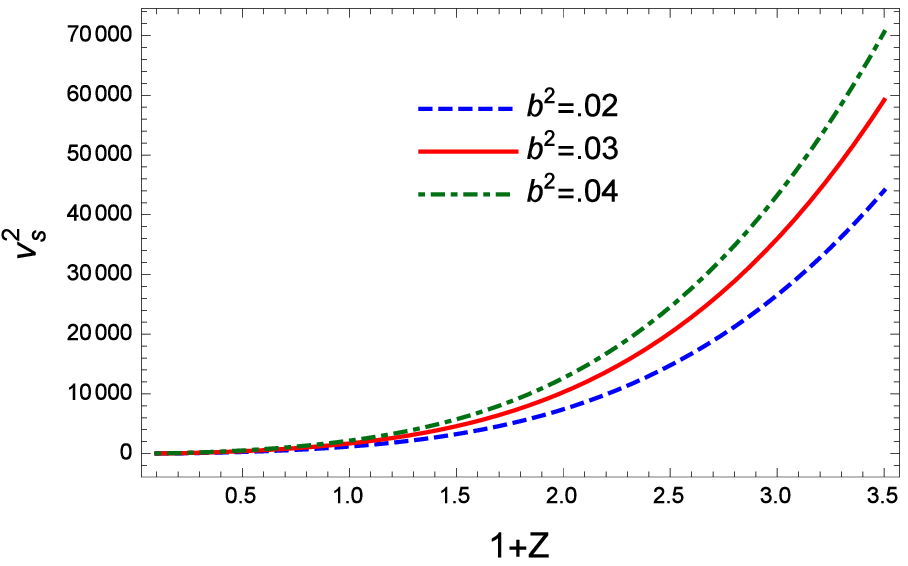}
\caption{Evolution of ${v}^{2}_{s}$ versus redshift parameter $z$ for
interacting NTADE. Here, we
have taken $\Omega^{0}_D=0.73$, $B=2.4$, $\delta=2.6$ and $H(a=1)=67$.
}\label{sn-z1}
\end{center}
\end{figure}

\begin{figure}[htp]
\begin{center}
\includegraphics[width=8cm]{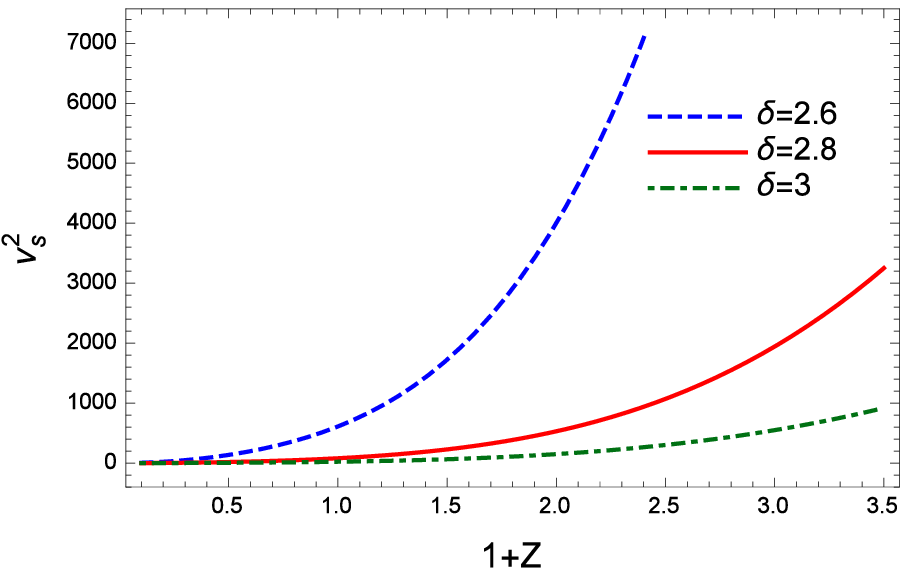}
\caption{Evolution of ${v}^{2}_{s}$ versus redshift parameter $z$ for
interacting NTADE. Here, we
have taken $\Omega^{0}_D=0.73$, $B=2.4$, $b^2=.01$ and $H(a=1)=67$.
}\label{ssn-z1}
\end{center}
\end{figure}

Considering the $Q=3b^2H(\rho_D+\rho_m)$ mutual interaction
between the dark sectors of cosmos \cite{int}, it is a matter of
calculations to find
\begin{eqnarray}\label{EoSna1}
&&\omega_D=-1-\frac{b^2}{\Omega_D}-\frac{2\delta-4}{3a\eta H},\nonumber\\
&&q=-\frac{1}{3}-\frac{3b^2}{2}-\frac{3\Omega_D}{2}-\frac{(\delta-2)\Omega_D}{a\eta H},\nonumber\\
&&\dot{\Omega}_D=\frac{(2\delta-4)\Omega_D}{a\eta}+2\Omega_D H(1+q),\nonumber\\
&&v_{s}^{2}=\frac{-7+3b^2+3\Omega_D}{6}\\&&-\frac{3^\frac{-7+4\delta}{-4+2\delta}b^2H(H^2\Omega_D B^{-1})^{\frac{1}{-4+2\delta}}
(-1+b^2+\Omega_D)}{6a(\delta-2)\Omega_D}\nonumber\\&&
-\frac{3^\frac{-3+2\delta}{4-2\delta}a(H^2\Omega_D B^{-1})^{\frac{1}{4-2\delta}}(5-2\delta+(\delta-2)\Omega_D)}{H},\nonumber
\end{eqnarray}
plotted in Figs.~\ref{wn-z1}-\ref{ssn-z1}. It is apparent that the
model is classically stable, and its parameters, including $q$,
$\omega_D$ and $\Omega_D$, have satisfactory behaviors. Thus, just
the same as the TADE model, the existence of the
$Q=3b^2H(\rho_D+\rho_m)$ interaction between the cosmos sectors
make the model stable at the classical level.
\section{Closing remarks}
Since gravity is a long range interaction, one should use the
generalized nonextensive Tsallis entropy for studying its related
phenomena
\cite{Tsallis1,TDE,THDE,THDE1,THDE2,THDE3,THDE4,THDE5,THDE6,SM,Ren}.
In this paper, inspired by the Tsallis entropy \cite{Tsallis1} and
based on the holographic hypothesis, we proposed a new DE model
with time scale as IR cutoff. We consider the age of the Universe
and the conformal time as system's IR cutoffs. The behavior of
$q$, $\omega_D$, $\Omega_D$ and ${v}^{2}_{s}$ have been studied
during the cosmic evolution. It was observed that in the absence
of interaction  both of these models are classically unstable. In
addition, we address the consequences of the existence of a mutual
interaction between the dark sectors of cosmos. We found out that,
unlike the original ADE models based on the Bekenstein-Hawking
entropy \cite{kyoung}, the interacting models introduced here are
classically stable. This is an interesting result which confirms
that interacting TADE and NTADE models may be useful in explaining the late time DE dominated universe. Our study shows also that the predictions of the models for the cosmic evolution are more sensitive to $\delta<1$ rather than $\delta>1$. This sensitivity is obtainable by comparing the corresponding curves, and is affected by the initial conditions used for plotting the curves. Holographic hypothesis is the backbone of the ADE models, a fact claiming that, in addition to the sing of the sound speed square, a full analysis on their stability should also consider their non-local features \cite{Stabil1,Stabil2}. The second approach is out of our goal in this paper, and can be considered as a serious issue for the future works.
\acknowledgments{We thank the Shiraz University Research Council. This
work has been supported financially by Research Institute for
Astronomy \& Astrophysics of Maragha (RIAAM), Iran.}

\end{document}